\documentclass{aastex}
\usepackage{spr-astr-addons}
\usepackage{url}
\usepackage{lscape, morefloats, graphicx, float}

\RequirePackage{color}

\begin{document}

\title{Gould Belt members in X-ray RAVE: Cross-matching RAVE stars with 3XMM point sources}
\shorttitle{Article preparation guidelines}
\shortauthors{B. Co\c skuno\u glu, O. Plevne, M. T. \" Ozkan}

\author{B. Co\c skuno\u glu \altaffilmark{1}}
\altaffiltext{1}{Istanbul University, Faculty of Science, Department 
of Astronomy and Space Sciences, 34119 University, Istanbul, Turkey\\
\email{basarc@istanbul.edu.tr}}

\author{O. Plevne \altaffilmark{2}} 
\altaffiltext{2}{Istanbul University, Graduate School of Science and Engineering, 
Department of Astronomy and Space Sciences, 34116, Beyaz\i t, Istanbul, Turkey\\}
\and
\author{M. T. \" Ozkan \altaffilmark{1,3}} 
\altaffiltext{3}{TUG (T\" UB\.ITAK National Observatory), Akdeniz University Campus, 07058, Antalya, Turkey\\}

\begin{abstract}
In this paper the results of matching the RAdial Velocity Experiment (RAVE), a spectroscopic Southern hemisphere survey (9 $ < I_{DENIS} <$ 12), and XMM-Newton Serendipitous Source Catalogue (3XMM) are presented. The latest data releases of RAVE and XMM were matched and a X-ray RAVE catalogue of 1071 stars was obtained. Then the catalogue was checked for possible Gould Belt (GB) members. We obtained a subsample of  10 stars that meet the GB membership criteria. This subsample and GB member candidates were tested photometrically and kinematically. Among the members there are two BY Dra type variables, an NGC2451 open cluster member, a high proper motion star. The rest are regular main sequence stars. The members have very low velocity dispersions which lead us to think that the members belong in a single structure. We also found out that a kinematical GB membership test might be possible to derive given a large enough GB member sample as they fit in a narrow interval in space velocity diagrams.
\end{abstract}

\section{Introduction}

\cite{Herschel47} observed that a fraction of bright stars in the southern hemisphere of the sky to be a part of a structure other than Milky Way with a tilt of $\approx$ 20$^\circ$ to the Galactic equator. 32 years later \cite{Gould79} also observed it and determined its poles' and nodes' Galactic coordinates. Therefore, this structure is named after him, namely the Gould Belt (hereafter GB). The GB is relatively flat: its length, width and height semiaxes are 350, 250 and 50 pc, respectively. Its ascending node is at $l$=285$^{\circ}$ and the Sun is 200 pc off the center in the $l$=130$^{\circ}$ direction \citep{Guillout98}. The GB consists of X-ray active young stars with distances up to 600 pc \citep{Olano82, Fresneau96, Poppel97, Guillout98}. The age interval varies greatly in the literature: most authors state that it is between 20-30 and 70-80 Myr \citep{Ogorodnikov65, Olano82, Poppel97, Guillout98, Bekki09}, however, \cite{Frogel77}'s review suggests a wider interval: 30-220 Myr. Using Str\" omgren photometry \cite{Torra00} determined stars with ages < 90 Myr belong in the GB. \cite{Bobylev14} wrote an extensive review regarding the GB, summarizing past studies about GB's formation, structure, age, kinematics, mass. The GB is important because it is the closest star-gas complex to the Sun. These complexes are regions where star formation is active and they are observed in Milky Way \citep{Efremov98a}, as well as other galaxies \citep{Efremov98b}.

The GB is not defined very precisely. There are several criteria that are used to define GB membership, which will be discussed in detail in Chapter 2. This causes controversial opinions to form about its formation. \cite{Bobylev14} states that the GB is a separate structure from the Milky Way, whereas \cite{Bekki09} says that a dark matter clump collided with a gas cloud to trigger GB's formation. \cite{Comeron01} observed a GB-like structure in M83 which is approximately the same size with the GB, is separate from the spiral structure of the galaxy, lies at a similar distance from the galactic center and has the same order of age ($\log$ age/yr=7). This means GB-like structures are not unique to our Galaxy and, at the very least, can exist in other spiral galaxies as well. Therefore, analysing and better defining the GB can help us understand the evolution of other galaxies. The aim of study is to determine GB members and their properties that are in both RAVE and 3XMM catalogues and to contribute towards defining the GB better. 

Having spectral information about stars in multiple regions of the electromagnetic spectrum is useful for many purposes in astronomy. We have cross-matched the RAdial Velocity Experiment (RAVE)'s catalogue with XMM-Newton Serendipitous Source Catalogue (3XMM) in order to obtain both infrared and X-ray data of the stars in common in the surveys in question. The resulting X-ray RAVE (XRAVE) catalogue was used to determine GB members. For the purposes of this study XMM-Newton data provides the X-ray luminosity of the stars, whereas the RAVE data provides the radial velocities, atmospheric parameters and distances of the stars in the matched catalogue. However, since the distances provided in RAVE have relatively large errors, the accurate distances of {\em GAIA} DR2, which includes almost all RAVE stars, were used \citep{Gaia18}. The atmospheric parameters of RAVE were utilised in determining ages using the method of \cite{Jorgensen05}.

The data and method are presented in Section 2, the properties of GB members are given in Section 3 and the conclusion and summary are in Section 4.

\section{Data \& Method}

\subsection{Data}

\begin{figure*}
\centering
\includegraphics[width=\textwidth]{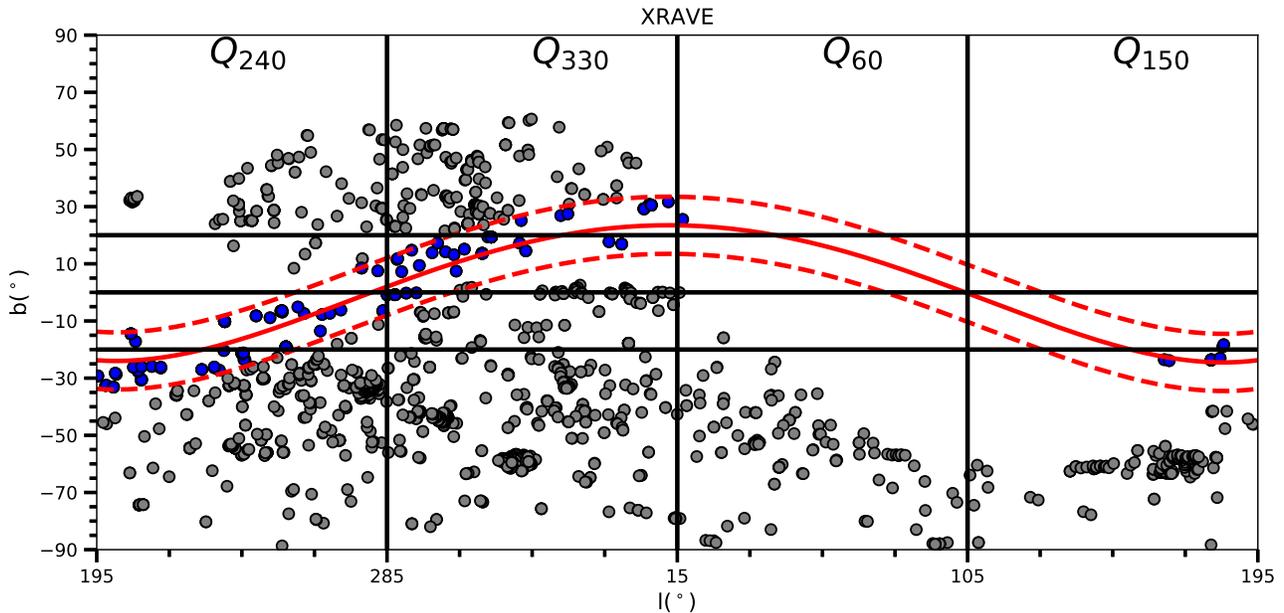}
\caption{92 XRAVE Stage 1 GB member candidates. Blue dots represent the candidates, gray dots represent XRAVE stars outside the GB region, the red solid line is the Galactic latitude obtained from Eq. 2, whereas the dashed red lines represent the thickness of the GB, i.e. GB's upper and lower edges.}
\end{figure*}

A part of the data used in creating the catalogue was taken from RAVE, which is a spectroscopic stellar survey based in the Southern hemisphere. RAVE's general aim is to make use of the Sun's position in the Galaxy and study its formation and evolution. As one can deduce from its name RAVE specializes in determining stellar radial velocities for stars in the thin and thick disks and halo. Its observations are performed with a 1.2 m Schmidt telescope in Anglo-Australian Observatory. Radial velocities are calculated using medium resolution spectra (R = 7500) in the Ca-triplet region (8400-8750 \AA) \citep{Steinmetz06}. Combining radial velocity with proper motion and distance one can obtain the kinematic components of a star. In the study we used RAVE's most recent data release, i.e. RAVE Data Release 5 (RAVE DR 5) \citep{Kunder17}. RAVE DR5 has 520781 spectra of 457588 unique stars \citet{Kunder17} and their radial velocities, atmospheric parameters, distances along with other parameters.

Another part of the data used in the study was taken from European Space Agency's XMM-Newton satellite's Serendipitous Source Catalogue's Data Release 8 (3XMM DR8) \citep{Rosen16}. XMM-Newton was launched in 1999. It has three X-ray telescopes, an X-ray detector with three CCDs, an optical monitor and two reflection grating spectrometers on board. Each telescope has 58 cylindrical, nested Wolter Type-1 mirrors which are 600 mm in length and 306 to 700 mm in diameter. European Photon Imaging Camera (EPIC), i.e. the X-ray detector has two Metal Oxide Semi-conductor (MOS) cameras and a p-n junction (PN) camera. The Optical Monitor (OM) helps XMM-Newton perform multi-wavelength observations simultaneously: in X-rays and optical and ultraviolet regions of the electromagnetic spectrum. 3XMM DR8 has 775153 detections of 531454 unique sources and provides X-ray fluxes in nine different bands as well as other parameters. 

We also used GAIA DR2 so that we could have higher astrometric accuracy. Launched in 2013, GAIA is the successor of RAVE and will observe stars brighter than 20$^m$. It aims to create the most accurate map of Milky Way. GAIA provides precise astrometric data, up to a few micro arc seconds of accuracy \citep{Gaia16}. 

Upon cross-matching the catalogues of both data releases and removing duplicates by keeping the highest signal-to-noise ratio observations and discarding others an XRAVE subsample of 1071 unique stars was obtained. Stars in RAVE DR5 and 3XMM DR8 catalogues were matched using the angular separation formula for equatorial coordinates (Eq. 1):

\begin{equation}
\theta=cos^{-1}[sin(\delta_1)sin(\delta_2)+cos(\delta_1)cos(\delta_2)cos(\alpha_1-\alpha_2)],
\end{equation}

where $\delta_1$ and $\delta_2$ are the declination of RAVE and XMM stars and $\alpha_1$ and $\alpha_2$ are the right ascension of RAVE and XMM stars, respectively. We accepted matches that yielded an angular separation of $\leq $ 1'' because of the high astrometic accuracy of RAVE \citep{Kunder17} and the increased astrometric accuracy of 3XMM \citep{Rosen16}. We checked our resulting catalogue using Vizier (\footnote{http://vizier.u-strasbg.fr/viz-bin/VizieR})'s X-match tool and obtained the same sample.

\subsection{Identification of GB members}

In order for a star to be a GB member it has to meet certain criteria for Galactic coordinates, distance, X-ray luminosity and age. These criteria are as follows:

\begin{itemize}
\item Galactic coordinates should fulfill Eq. 2,
\item Distances should be less than 600 pc,
\item X-ray luminosities in 0.2-12 keV energy band should be between 10$^{28}$ and 10$^{31}$ ${\rm erg\,s^{-1}}$,
\item Ages should be between 10 and 100 Myr.
\end{itemize}

In order to identify the stars that meet the first criterion we used the GB description given by \cite{Guillout98} and shown in their Fig. 4. Their description can be best represented with a fourth degree polynomial that associates Galactic latitude with Galactic longitude (Eq. 2).  

\begin{eqnarray}
b & =5.50\times10^{-8} \times l^4-8.19 \times 10^{-5} \times l^3 \nonumber \\
   & +4.25\times10^{-2} \times l^2 - 8.96 \times l + 634,
\end{eqnarray}

where $l$ and $b$ are Galactic longitude and latitude, respectively. A fourth degree polynomial was used because it most accurately describes GB's position in Galactic coordinates. Higher degree polynomials yield almost the same amount of stars with very insignificant coefficients in the order of 10$^{-10}$, which have no physical meaning, whereas lower degrees distort and flatten the shape of GB. Considering the GB has a thickness of $\approx$ 20$^\circ$ \citep{Comeron94} in latitude, we accepted stars within $\pm$ 10$^{\circ}$ of the latitude provided by Eq. 2 for the longitude in question. The GB we obtained from Eq. 2 is given in Fig. 1. 92 out of 1071 XRAVE stars fall within the dashed red lines in Fig. 1, i.e. they are within the GB sky area. We refer to these stars as Stage 1 GB candidates, since they passed one criterion out of four.   

After sorting out Galactic positions of GB candidates, we moved on to distances. Since the distances given in RAVE DR5 have average relative errors of greater than 40 per cent, we used GAIA DR2's parallaxes. The average relative error of parallaxes in GAIA is $\approx$ 5 per cent. Therefore, we adopted GAIA parallaxes in order to determine distances and X-ray luminosities using those distances. One star in XRAVE had no match in GAIA, therefore it was excluded, reducing our Stage 1 subsample to 91 stars. After removing three stars with zero parallax and 32 stars with distances larger than 600 pc we ended up with 56 Stage 2 XRAVE stars that meet both the position and distance criteria for GB membership (Fig. 2).

\begin{figure}[h]
\centering
\includegraphics[width=\columnwidth]{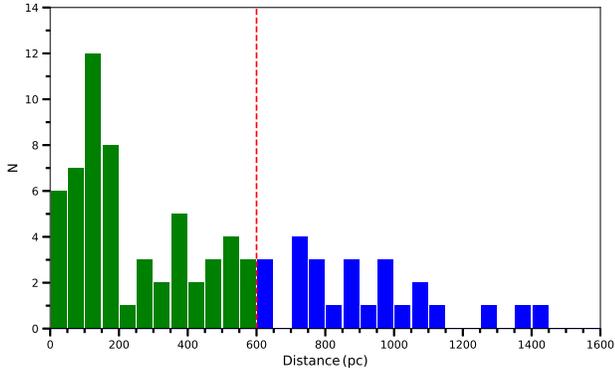}
\caption{Distance histogram of 91 Stage 1 GB candidates. The red dashed line represents the distance limit for GB, namely 600 pc. The stars in green histograms, whose distances are below the limit, i.e. stars that remain to the left of the dashed line are the Stage 2 candidates, stars to the right of the dashed line in blue histograms remain as Stage 1 candidates.}
\end{figure}

In the next step we determined X-ray luminosities of our Stage 2 candidates. \cite{Guillout98, Stelzer00} stated that $L_X$ shows a broad distribution between 10$^{28}$ and 10$^{31}$ erg.s$^{-1}$ for young stars which is the interval we adopted for our GB candidates. After calculating the distances and using them to compute luminosities along with the fluxes from the 0.2-12 keV band. The 0.2-12 keV band, which is the widest energy band in XMM catalogue, was adopted in order to obtain matches from all types of stars. The luminosities were computed as follows:

\begin{equation}
L_X=4\pi d^2F_{0.2-12},
\end{equation}

where $d$ is the distance, $F_{0.2-12}$ is the X-ray flux of the star in the 0.2-12 keV energy band. Afterwards, we looked for the stars with X-ray luminosities in the 10$^{28}$ and 10$^{31}$ erg.s$^{-1}$ interval \citep{Guillout98}. We obtained a sample of 49 Stage 3 stars meeting all three criteria, i.e. Galactic position, distance and X-ray luminosity. The X-ray luminosity distribution of 56 Stage 2 candidates are shown in Fig. 3.

\begin{figure}[h]
\centering
\includegraphics[width=\columnwidth]{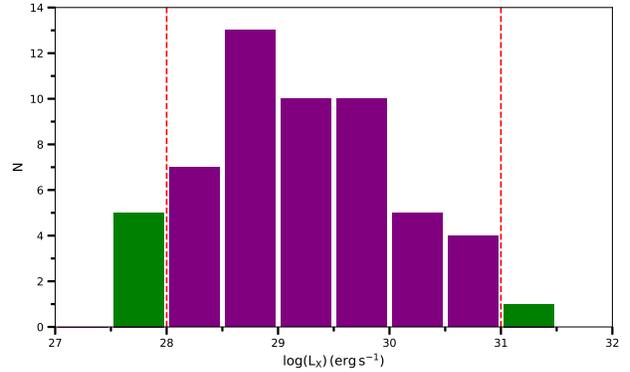}
\caption{X-ray luminosity distribution of 56 Stage 2 GB candidates. The dashed red line to the left and right represent X-ray luminosities of 10$^{28}$ and 10$^{31}$ erg.s$^{-1}$, respectively. The stars in purple histograms that are between the lines are the 49 Stage 3 candidates, the stars in green histograms remain as Stage 2 candidates.}
\end{figure}

\subsection{Calculating ages of GB stars}

The last step towards determining GB members is calculating ages. Since neither RAVE and XMM catalogue nor GAIA provide ages we used the Bayesian method of \cite{Jorgensen05} in order to derive ages. This method includes probability density functions (PDF) that are obtained from  theoretical model parameters and observational parameters \citep{Pont04, Jorgensen05}. \citet{Pont04} stated that the Solar neighborhood mass, age and metallicity prior distributions are assumed to be independent from each other, hence they are taken into account separately. \cite{Jorgensen05} assumed that observational parameter errors are independent and that they show a normal distribution. According to these assumptions the likelihood function is defined as:

\begin{equation}
{\cal L}(\tau, \zeta, m) = \prod_{i=1} ^{n} \frac{1}{\sqrt{2 \pi} \sigma_{i}} \exp{[-\chi ^2 /2]}.
\label{eq:likelihood}
\end{equation}
$\chi ^2$ is given as
\begin{equation}
\chi ^2 = \sum_{i=1}^{n} \left( \frac{q_{i}^{obs} - q_{i}(\tau , \zeta , m) }{ \sigma_i} \right),
\label{eq:chi2}
\end{equation}

where $q$ represents the model atmospheric parameters taken from RAVE DR5, i.e. $T_{eff}$, $\log g$ and $[Fe/H]$, whereas $\tau$, $\zeta$ and $m$ are the theoretical model parameters which represent age, metallicity, and mass, respectively. $n$ is the number of objects and $\sigma$ represent parameter errors. Then, the final PDF is obtained by applying a Bayesian correction given below:

\begin{equation}
f (\tau , \zeta , m) \varpropto f_0  (\tau, \zeta, m) \times {\cal L}(\tau, \zeta, m)
\label{eq:bayesian}
\end{equation}

where ${f_0}$ is the initial pdf and it is given as

\begin{equation}
f_0=\psi(\tau) \varphi (\zeta | \tau) \xi(m | \zeta , \tau).
\label{eq:priors}
\end{equation}

In this equation $\psi (\tau)$, $\varphi (\zeta)$ and $\xi (m)$ are the star formation rate (SFR), the metallicity distribution (MDF) and the initial mass function (IMF), respectively. SFR and MDF are assumed to be constant since $\tau$, $\zeta$ and $m$ are independent from each other. However, since there are more low mass stars than high mass stars in the Galaxy, the IMF is expressed as a power law \citep{Jorgensen05}. We then inserted Eq. (7) into Eq. (6) and integrated the resulting equation and obtained the $G(\tau)$ function (Eq. 8) which shows the PDF calculated from a comparison between observational and theoretical parameters.

\begin{equation}
G(\tau) \varpropto \int \int {\cal L}(\tau, \zeta, m) \xi (m) dm d\zeta
\label{eq:Gfunction}
\end{equation}

We integrated Eq. 8 over mass and metallicity for each age on isochrones. Taking into account the maximum value of the $G$ function we determined the age for the star in question. Due to the complex and non-Gaussian nature of Bayesian age estimation method, overall uncertainties on estimated star ages vary between 20 per cent and 50 per cent. 

We applied this method to Stage 3 candidates using the PARSEC \citep{Bressan12} stellar evolution models. Since GB members are young stars with ages between 30 and 80 Myr \citep{Fresneau96, Guillout98}, we adopted 10 Myr to be the lower limit to eliminate Classical T Tauri Stars \citep{Bobylev14} and 100 Myr \citep{Wichmann03} to be the upper limit for this study to make up for errors. The age distribution of XRAVE is shown in Fig. 4.  

\begin{figure}[h]
\centering
\includegraphics[width=\columnwidth]{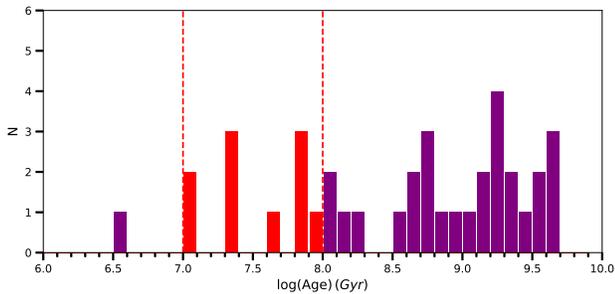}
\caption{Age distribution of 49 Stage 3 GB candidates. The dashed red line to the left represents 10 Myr, whereas the one to the right stands for 100 Myr, which are the adopted lower and upper limits for GB members for this study, respectively. 10 stars between the lines that are in the red histograms are the members, and the ones in purple histograms that are not between the lines remain Stage 3 candidates.}
\end{figure}

As seen from the figure, there are 10 GB members. There is also a star with age 3 Myr that did not meet the lower limit of GB membership criterion even within error bars. After checking the literature for this star we have found that it is an Algol type eclipsing binary \citep{Avvakumova13}, which means it would throw off RAVE measurements yielding such an unpredictable and possibly incorrect age. This star's age was calculated as $\approx$ 3 Myr and is the only star below the age limit as seen in Fig. 4.

Fig. 5 shows GB candidates for all stages: Stage 1, Stage 2, Stage 3 and member. 35 Stage 1 candidates which only passed the Galactic position criterion are represented by blue circles, 7 Stage 2 candidates which passed the Galactic position and distance criteria are denoted by green circles, 39 Stage 3 candidates which passed Galactic position, distance and X-ray luminosity criteria are shown with purple circles and finally 10 members who meet all criteria are given in big red circles.

\begin{figure*}[h]
\centering
\includegraphics[width=\textwidth]{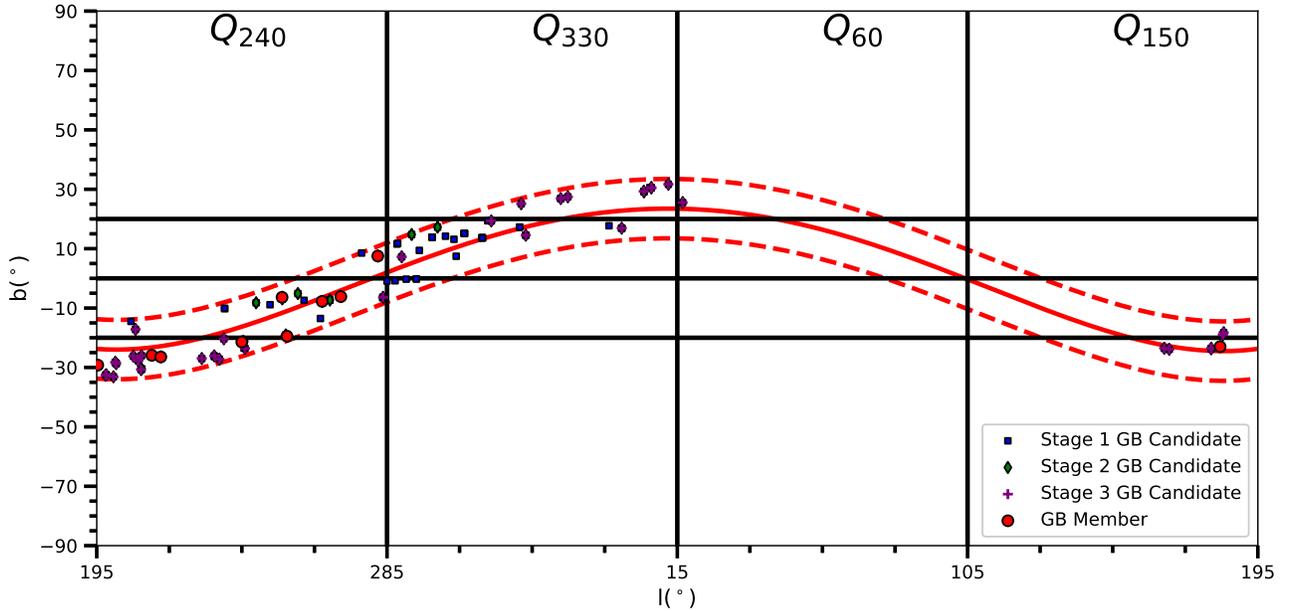}
\caption{Galactic distribution of all 91 GB member candidates. Blue dots represent Stage 1 GB candidates, green dots stand for Stage 2, whereas purple dots are Stage 3 and finally, the red ones represent GB members.}
\end{figure*}

\section{Properties of GB members}

In this section we discuss the 10 stars, which meet all four aforementioned GB membership criteria, i.e. GB members. Their Galactic coordinates, RAVE and XMM catalogue identifiers, distances, X-ray luminosities in the 0.2-12 keV energy band and ages and the errors for these three parameters are given in Table 1. 

\begin{table*}[h]
\caption{Galactic positions, catalogue identifiers, distances, distance errors, X-ray luminosities in 0.2-12 keV energy band, luminosity errors, logarithmic ages and age errors of GB members.}
\scriptsize
\begin{tabular}{cccccccccc}
\hline
$l$ & $b$ & RAVEID & XMMID & $d$ & $\sigma_d$ & $L_{0.2-12}$ & $\sigma_{L_{0.2-12}}$ & $\log(age)$ & $\sigma_{\log(age)}$ \\
\hline
$^\circ$ & $^\circ$ & & & (pc) & (pc) & $\times 10^{30} {\rm erg\,s^{-1}}$ & $\times 10^{30} {\rm erg\,s^{-1}}$ & (yr) & (yr) \\
\hline
183.3054 & -22.9945 & J043225.7+130648 & 3XMM J043225.7+130645 & 46.8 & 0.1 & 0.45 & 0.02 & 7.40 & 0.06 \\
195.3055 & -29.2149 & J043636.6+004331 & 3XMM J043636.5+004330 & 141.2 & 0.7 & 0.45 & 0.07 & 7.39 & 0.05 \\
212.0724 & -25.9166 & J051633.7-104740 & 3XMM J051633.7-104735 & 133.1 & 0.5 & 0.27 & 0.11 & 7.99 & 0.75 \\
214.8583 & -26.4542 & J051848.1-131732 & 3XMM J051848.0-131731 & 465.5 & 7.0 & 2.22 & 0.29 & 7.87 & 1.34 \\
239.9732 & -21.4176 & J061515.4-325403 & 3XMM J061515.6-325403 & 151.0 & 0.7 & 0.02 & 0.01 & 7.08 & 0.02 \\
253.9847 & -19.4802 & J064646.5-444036 & 3XMM J064646.6-444036 & 440.0 & 113.8 & 2.27 & 0.93 & 7.89 & 0.11 \\
252.4563 & -6.3732 & J074700.5-375037 & 3XMM J074700.3-375036 & 178.2 & 7.1 & 0.07 & 0.05 & 7.90 & 0.26 \\
264.8783 & -7.7405 & J081554.9-490555 & 3XMM J081554.9-490554 & 363.5 & 2.9 & 8.26 & 0.15 & 7.38 & 0.01 \\
270.6851 & -6.1511 & J084526.9-525202 & 3XMM J084527.0-525201 & 150.3 & 0.5 & 0.71 & 0.02 & 7.66 & 0.01 \\
282.1352 & 7.5193 & J103851.9-495522 & 3XMM J103851.7-495521 & 597.4 & 10.7 & 1.25 & 0.29 & 7.06 & 0.01 \\
\hline
\end{tabular}
\end{table*}

\subsection{Evolutionary status of GB members}

The Hertzsprung-Russell (HR) diagram for the entire XRAVE catalogue (1071 stars) is given Fig. 6. The bands used in the axes are from 2MASS \citep{Cutri03} photometry because that is readily provided by RAVE. The de-reddening procedure of $J$ and $H$ bands are as follows: we used \cite{Schlegel98}'s maps and obtained $E_\infty(B-V)$. Then we reduced $E_\infty(B-V)$ to $E_{d}(B-V)$ by taking into account distances obtained from the parallaxes provided by GAIA DR2 as follows: 

\begin{equation}
E_d(B-V)=E_\infty(B-V) \Biggl[1-\exp\Biggl(\frac{-\mid d~\sin(b)\mid}{H}\Biggr)\Biggr],
\end{equation}

where $b$ is the Galactic latitude and $d$ is the distance of the star, whereas $H$ is the scaleheight for the interstellar dust which is 125 pc \citep{Marshall06}. $E_{\infty}(B-V)$ stands for the maximum theoretical colour excess at the Galaxy's edge for that direction and $E_d(B-V)$ is the reduced absorption adjusted for distance in that direction. After obtaining $E_d(B-V)$ we calculated the total reddening for $J$ and $H$ bands using \cite{Fiorucci03}'s equations:

\begin{eqnarray}
A_J=0.885\times E_d(B-V), \nonumber \\
A_H=0.565\times E_d(B-V),
\end{eqnarray}

where $A_J$ and $A_H$ are the distance adjusted absorptions in $J$ and $H$ bands in the direction of the star in question. The absorptions were used in Pogson's equation to determine absolute magnitudes:

\begin{eqnarray}
m_J-M_J=5\log (d/pc)-5+A_J, \nonumber \\
m_H-M_H=5\log (d/pc)-5+A_H.
\end{eqnarray}

After de-reddening the 2MASS photometric bands and calculating absolute magnitudes for all XRAVE stars, we plotted the HR diagram (Fig. 6). There are several giants as seen in the figure, mostly confirmed by their RAVE surface gravities. The main sequence can be clearly observed, and the number of stars that belong there comprise the majority of the sample. This is expected because of RAVE's observation criteria as it focuses mainly on relatively bright main sequence stars with F or G spectral types. The turnoff point is also apparent in the figure. The 10 GB members are also mostly main sequence as well with a few giants. 

\begin{figure}[h]
\centering
\includegraphics[width=\columnwidth]{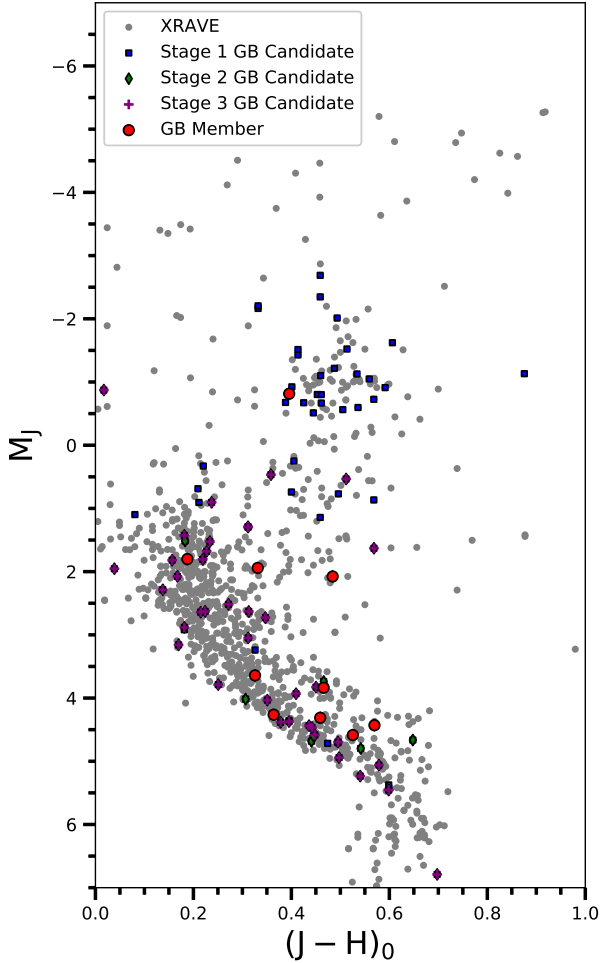}
\caption{HR diagram for the entire XRAVE sample. Grey circles are non GB stars, blue circles denote Stage 1 candidates, green circles are for Stage 2, whereas purple circles represent Stage 3 and big red circles are the GB members. The x-axis is the de-reddened $(J-H)_0$ colour, the y-axis is the absolute $J$ magnitude.}
\end{figure}

\subsection{Special GB members}

There are interesting objects amongst the GB members. We provide their atmospheric parameters, i.e. effective temperature, surface gravity and metallicity along with their $J$, $H$ and $K_S$ magnitudes in Table 2. The "Remarks" column provides information about a star if it has any peculiar features. Two of the members are BY Dra variables, which are chromospherically active late spectral type main sequence stars. Their effective temperatures and surface gravities confirm this result. One member is a high proper motion star. Another is a member of the open cluster NGC2451, which has an age of 30-50 Myr \citep{Platais01, Hunsch03}, which is in agreement with our estimate of 80 Myr. This result lends credence to our ages. However, it is unexpected to have a single member of an open cluster in the GB, therefore, we checked XRAVE for other NGC2451 members. We looked for its members in WEBDA \footnote{https://www.univie.ac.at/webda/navigation.html} and found 136 of them \citep{Williams67}. Among those 136 stars only the one in the GB is in XRAVE, and seven are in RAVE DR5. This means the other six stars in RAVE DR5 were either not observed by XMM-Newton which is likely because the GB member is at the edge of the open cluster or are not X-ray active. As a result, we could not check if other NGC2451 members belong in the GB or not. However, according to \cite{Elias09}'s spatial criteria method more than half of the members of NGC2451 are part of the GB. This suggests that we could only find a single member of NGC2451 in the GB, because we did not have membership criteria related data for its other members. Two of the GB members are well past their turn off points and are giants, more specifically, one of them is a subgiant and the other is a giant. Their surface temperatures suggest that they are G-type stars. Taking this into account along with their young ages, it does not seem possible. The reason for this discrepancy might be faulty atmospheric parameter determination due to the RAVE pipeline as it is adjusted for main sequence stars. A literature check has not yielded any decisive results, neither for the two stars to be giants, nor against it.

\begin{table*}[h]
\caption{Atmospheric parameters and 2MASS photometry bands' magnitudes of GB members.}
\begin{tabular}{ccccccccc}
\hline
$l$ & $b$ & $T_{eff}$ & $\log\,g$ & $[Fe/H]$ & $M_J$ & $(J-H)_0$ & $(H-K_s)_0$ & Remarks\\
$(^\circ)$ & $(^\circ)$ & (K) & $(cm\,s^{-2})$ & (dex) & & & &\\
\hline
183.3054 & -22.9945 & 3943 & 4.97 & -0.31 & 4.58 & 0.53 & -0.03 & BY Dra variable\\
195.3055 & -29.2149 & 4000 & 5.00 & 0.00 & 4.43 & 0.57 & 0.13 &\\
212.0724 & -25.9166 & 5030 & 4.88 & 0.00 & 4.26 & 0.36 & 0.08 & high proper motion\\
214.8583 & -26.4542 & 7262 & 3.95 & 0.00 & 1.80 & 0.19 & 0.06 &\\
239.9732 & -21.4176 & 4932 & 4.38 & 0.39 & 3.84 & 0.47 & 0.08 &\\
253.9847 & -19.4802 & 5000 & 4.50 & 0.00 & 2.08 & 0.48 & 0.12 &\\
252.4563 & -6.3732 & 5025 & 4.90 & 0.10 & 3.64 & 0.33 & -0.14 & member of NGC 2451\\
264.8783 & -7.7405 & 5678 & 1.25 & 0.00 & 1.94 & 0.33 & -0.06 &\\
270.6851 & -6.1511 & 4000 & 4.99 & 0.00 & 4.31 & 0.46 & -0.09 & BY Dra variable\\
282.1352 & 7.5193 & 4925 & 2.03 & -0.61 & -0.81 & 0.40 & -0.03 &\\
\hline
\end{tabular}
\end{table*}

\section{Conclusion and Summary}

In this study we cross matched RAVE DR5 and 3XMM DR8 and obtained a resulting XRAVE catalogue of 1071 stars. We looked for GB members in XRAVE. According to \cite{Olano82, Poppel97, Guillout98, Torra00} a member of GB has to meet four criteria as discussed in Section 2: a certain Galactic position, distance less than 600 pc, X-ray luminosity between 10$^{28}$ and 10$^{31}$ ${\rm erg\,s^{-1}}$ and age between 10 and 100 Myr. We applied those criteria to XRAVE and obtained a sample of 10 stars that meet all of them, i.e. that are GB members. 

In order to test the GB membership criteria with another method we approached the GB candidate sample of 91 stars kinematically. $U$, $V$ and $W$ Galactic velocity components were obtained using distances and radial velocities in conjunction with radial and tangential proper motions in \cite{Johnson87}'s algorithms and transformation matrices. A detailed explanation of the calculations is given by \cite{Coskunoglu11, Coskunoglu12}. After obtaining total space velocities Galactic differential rotation corrections were obtained and performed for the $U$ and $V$ components using \cite{Mihalas81}'s procedure. The $W$ component is not affected by Galactic differential rotation \citep{Mihalas81}. Then we performed the Local Standard of Rest (LSR) correction on the Galactic velocity components as provided by \cite{Coskunoglu11}. LSR correction is required to determine velocities accurately as it describes the Sun's movement with respect to nearby stars \citep{Coskunoglu11}. Finally, the errors of Galactic velocity components were calculated \cite{Johnson87}'s algorithm by propagating the errors of distances, radial velocities and radial and tangential proper motions. The median space velocity components and median total space velocity and their respective errors along with the velocity dispersions for each component and total velocity are given in Table 3 for Stages 1, 2 and 3 and GB members. The values in the table are in km s$^{-1}$.

\begin{table*}[h]
\caption{Median space velocities and their respective errors and dispersions of 91 Stage 1, 56 Stage 2, 49 Stage 3 candidates and 10 members. All velocities are given in km s$^{-1}$.}
\begin{tabular}{c|cccccccccccc}
\hline
&$\tilde{U_{LSR}}$ & $\tilde{V_{LSR}}$ & $\tilde{W_{LSR}}$ & $\tilde{S_{LSR}}$ & $\tilde{U_{Err}}$ & $\tilde{V_{Err}}$ & $\tilde{W_{Err}}$ & $\tilde{S_{Err}}$ & $\sigma_{U_{LSR}}$ & $\sigma_{V_{LSR}}$ & $\sigma_{W_{LSR}}$ & $\sigma_{S_{LSR}}$ \\
\hline
Stage 1&-29.81 & -30.26 & -14.14 & 57.55 & 1.46 & 1.41 & 0.82 & 2.48 & 227.64 & 306.03 & 492.95 & 605.07 \\
Stage 2&-31.26 & -30.58 & -14.31 & 54.33 & 1.20 & 1.11 & 0.75 & 2.32 & 82.71 & 84.39 & 41.13 & 108.17 \\
Stage 3&-22.50 & -32.44 & -14.83 & 48.90 & 1.10 & 1.00 & 0.71 & 2.16 & 100.16 & 33.01 & 49.41 & 107.41 \\
GB &-18.44 & -34.94 & -11.36 & 47.33 & 1.12 & 1.85 & 0.78 & 3.20 & 24.56 & 4.65 & 14.26 & 10.78 \\
\hline
\end{tabular}
\end{table*}

Upon analysing Table 3 it can be seen that every GB criterion that was applied to XRAVE catalogue also applies a selection effect kinematically, which can be observed from the decreasing space velocity dispersions in every step. The GB members' low space velocity dispersions and their low errors show that those stars have similar kinematic characteristics. In order to demonstrate that we plotted a Toomre diagram, which shows the kinematical similarities or the lack thereof for star samples, of the entire XRAVE sample (Fig. 7). Every step reduces the scatter of stars in the diagram noticeably and ultimately yields a very narrow interval for the GB members (red circles in Fig. 7), which verifies the dispersion decrease in Table 3. This points out that GB can be detected kinemetically as well, however, the lack of number of stars in the sample prevents the determination of a distinguishable feature. This can be countered by working with larger samples and thus obtaining bigger samples of GB members. After that is done kinematical constraints can be determined for the GB. 

\begin{figure}[h]
\caption{Toomre diagram for the entire XRAVE sample. Grey circles are non GB stars, blue circles denote Stage 1 candidates, green circles are for Stage 2, whereas purple circles represent Stage 3 and big red circles are the GB members. The x-axis is the $V_{LSR}$, the y-axis is $\sqrt{U_{LSR}^2+W_{LSR}^2}$ both in km s$^{-1}$.}
\centering
\includegraphics[width=\columnwidth]{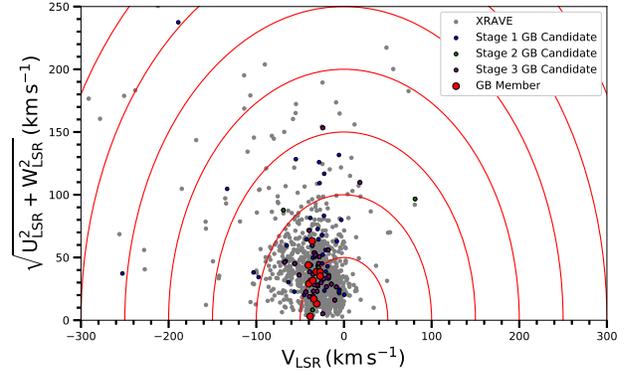}
\end{figure}

Our analyses show that there are 10 GB members which appear in both RAVE DR 5 and 3XMM DR 8 catalogues. We tested our subsample photometrically and kinematically. Our findings indicate that the 10 stars are part of a single structure. The majority of the members are regular F or G spectral type main sequence stars as expected from RAVE's selection criteria. There are two BY Dra type variables which are chromospherically active stars. A NGC2451 open cluster member and two giants also belong in the member subsample. As giants of F or G spectral type need to be at least billions of years old it is possible that faulty atmospheric parameter detection threw off the calculations. We also found that a kinematical GB membership test can be derived with a large enough GB member sample. If such a test would exist it would help in defining the GB better and more precisely. \cite{Comeron01} also remarked that the GB's inclination to the Galactic plane might show up as a systematic pattern in the radial velocities of the HII regions, which would help in deriving the aforementioned kinematical membership test. In order to obtain such a sample larger spectroscopic surveys such as Mauna Kea Spectroscopic Explorer (MSE) used in conjunction with future data releases of GAIA are required.

\section*{Acknowledgements}

We would like to thank Dr. Sinan Ali\c s for his numerous contributions towards improving the paper. Also, we thank Prof. A. T. Sayga\c c and Dr. G. G\" un for their input towards developing the study.

\end{document}